\documentclass[aps,twocolumn]{revtex4}
\usepackage{amssymb}
\usepackage{amsmath}
\usepackage{amsfonts}
\usepackage{epsfig}
\usepackage{bbm}
\usepackage{comment}
\usepackage{multirow}
\usepackage{array}
\usepackage{subfigure}
\usepackage{graphics}
\usepackage{xcolor}
\usepackage{graphicx} 
\usepackage{braket}
\usepackage{rotating}
\usepackage{float}
\usepackage{mwe}    % loads �blindtext� and �graphicxï¿Ref.{½
\usepackage{subfig}
\usepackage{color}
\usepackage{empheq}
\begin{document}

\title{Localization of pairing correlations in nuclei within  relativistic mean field models}
\author{R.-D. Lasseri}
\affiliation{Institut de Physique Nucl\'eaire, Universit\'e Paris-Sud, IN2P3-CNRS, 
F-91406 Orsay Cedex, France}
\author{J.-P. Ebran}
\affiliation{CEA,DAM,DIF, F-91297 Arpajon, France}
\author{E. Khan}
\affiliation{Institut de Physique Nucl\'eaire, Universit\'e Paris-Sud, IN2P3-CNRS, 
F-91406 Orsay Cedex, France}
\author{N. Sandulescu}
\affiliation{National Institute of Physics and Nuclear Engineering, P.O. Box MG-6, 077125 Magurele,
Romania}

\begin{abstract}
We analyze the localization properties of two-body correlations induced by pairing in the framework of relativistic
mean field (RMF) models. The spatial properties of two-body correlations are studied for the pairing tensor in coordinate
space and for the Cooper pair wave function. The calculations are performed both with Relativistic-Hatree-Bogoliubov
(RHB) and RMF+Projected-BCS (PBCS) models and taking as examples the nuclei $^{66}$Ni, $^{124}$Sn and $^{200}$Pb.
It is shown that the coherence length have the same pattern as in previous non-relativistic HFB calculations, i.e.,
it is maximum in the interior of the nucleus and drops to a minimum in the surface region. In the framework of
RMF+PBCS we have also analysed, for the particular case of $^{120}$Sn, the dependence of the coherence length
on the intensity of the pairing force. This analysis indicates that  pairing is reducing the coherence length 
by about 25-30 $\%$ compared to the RMF limit.

\end{abstract}

\maketitle

\section{Introduction}
Pairing properties  of Fermi systems are commonly described in BCS-type approximations. In this framework the pairing correlations are 
characterized by global quantities such as  pairing gap and pairing energy. In finite systems such as atomic nuclei the  pairing 
properties  are strongly influenced by the density of levels  close to the chemical potential, which depends significantly on 
the number of nucleons and the spin-orbit interaction.  A peculiarity of finite systems 
is that the pairing correlations have a non-uniform spatial distribution. The latter is usually described
by the pairing tensor in coordinate space. For atomic nuclei  the localization properties of the pairing
tensor have been  studied in the non-relativistic Hartree-Fock Bogoliubov  (HFB) approach 
\cite{matsuo,pillet07,pillet10,vinas,pastore}.  As expected,  the
HFB calculations have shown that in nuclei the spatial distribution of pairing correlations  depends strongly
on the localization properties of  single-particle states  from the vicinity  of Fermi level. Less expected was
however the behavior of the coherence length which has emerged from the HFB calculations.
It was found that in nuclei the coherence length has a generic pattern: it is maximum close to the center of 
the nucleus and drops to a minimum  of 2-3 fm far out in the surface region. Similar short range two-body
correlations have been noticed also in semi-infinite neutron matter \cite{baldo}, in nuclei with two
neutrons extra a  closed core \cite{ibarra,catara} and in two-neutron halo systems such as $^{11}$Li
\cite{hagino}.

The scope of this paper is to study how the localization properties of pairing correlations
mentioned above  are affected by: (1)  the level density and the spin-orbit interaction associated to a relativistic 
mean field (RMF); (2) the restoration of the particle number conservation. These two issues will be discussed
in the framework of Relativistic-Hartree-Bogoliubov (RHB)  and RMF+Projected BCS (PBCS) models.  

In BCS-type models the active fermions participating to pairing correlations are described by antisymmetrised 
products of  two-body  functions, the  Cooper pairs, which are the same for any pair of  fermions.  
This is the basic assumption of BCS-type models from which the superfluid and superconducting 
properties are emerging. The size of the Cooper pairs is essential for characterizing the
pairing regime of the system. Thus, if the dimension of the Cooper pairs are larger or smaller 
than the mean distance between the fermions, the system is, respectively, in the BCS or Bose-Einstein 
condensation (BEC) regime. According to our knowledge, there are no systematic studies on the 
localization properties of Cooper pairs in nuclei. Thus, another scope of this paper is to present 
such a study in the framework of the two approximations mentioned above, RHB and RMF+PBCS.

\section{Formalism and calculation scheme}\label{rmf}
The two-body correlations of superfluid type in  many-body systems are commonly expressed by the reduced 
two-body density \cite{yang}. Of interest here is the diagonal part of the two-body density in 
coordinate space. 
Its expression for  one type of fermions, e.g., neutrons
or protons, is given by
\begin{equation}
\rho^{(2)} (x_1,x_2) =\bra{\Psi} \hat{\psi}^\dagger(x_1)\hat{\psi}^\dagger(x_2)\hat{\psi}(x_2)\hat{\psi}(x_1)\ket{\Psi},
\label{tbd}
\end{equation} 
where  $\ket{\Psi}$ is the ground state of the system and $\hat{\psi}^\dagger(x)$ is the operator which creates a nucleon with the
space-spin coordinates $x \equiv (\boldsymbol{r},\sigma)$.
The two-body density (\ref{tbd}) gives the probability to find in the system two nucleons with the coordinates $x_1$ and $x_2$.

The two-body correlations induced by pairing are usually studied in the BCS-like models in which
the ground state is approximated by
\begin{equation} 
\ket{BCS} = \prod_k\left(u_k+v_ka^\dagger_k a^\dagger_{\bar{k}}\right)\ket{0}
                 \propto \sum_n \frac{(\Gamma^+)^n}{n!},
\label{bcs}                 
\end{equation}
where $a^\dagger_k$ creates a particle in the single-particle state k and $u_k$ and $v_k$ are the standard 
variational parameters of the BCS theory.  The operator   $\Gamma^+\equiv\sum_k \frac{v_k}{u_k} a^\dagger_k a^\dagger_{\bar{k}} $
creates a  collective Cooper pair by scattering two nucleons in time reversed states $(k,\bar{k})$.  

In the BCS approximation the two-body density matrix has the expression    
\begin{equation}
\rho^{(2)}_{BCS} (x_1,x_2) = \rho^{(1)}(x_1)\rho^{(1)}(x_2)+ |\kappa(x_1,x_2)|^2 ,
\label{dev}
\end{equation}
where $\rho^{(1)}$ and $\kappa$ are the one-body density and the pairing tensor in the
coordinate representation defined by
\begin{empheq}{align}
&\rho^{(1)}(x) \equiv \bra{BCS}\hat{\psi}^\dagger (x) \hat{\psi}(x)\ket{BCS}, \\
&\kappa (x_1,x_2) \equiv \bra{BCS}\hat{\psi}(x_2)\hat{\psi}(x_1)\ket{BCS}.
\end{empheq} 
In a single-particle basis  the pairing tensor can be written as
\begin{equation}
\kappa(x_1,x_2)= \sum_k u_k v_k f_k(x_1) f_{\bar{k}}(x_2) , 
\end{equation} 
where $f_k(x)$ are the single-particle wave functions. This expression is formally the same for the pairing models
based on a general Bogoliubov transformation, such as HFB and RHB models. In this case the parameters
$u_k$ and $v_k$ and the single-particle wave functions $f_k(x)$  correspond to the canonical basis 
(e.g., see \cite{ring_schuck}).

According to Eq. (\ref{dev}), the quantity $|\kappa (x_1,x_2)|^2$  accounts for the genuine two-body
correlations in spin and spatial coordinates which are not included in the first mean-field term.
Since, by definition, $\kappa(x_1,x_2)$ is related  to the two-particle transfer amplitude, 
the corresponding two-body correlations can be probed by two-particle transfer reactions.

For the study of two-body pairing correlations of special interest is the quantity $\kappa (\boldsymbol{r_1},\sigma;\boldsymbol{r_2},-\sigma)$.
This quantity  gives the probability to find in the system two nucleons with opposite spins and located 
at $\boldsymbol{r_1} $ and $\boldsymbol{r_2}$. The spatial correlations are usually expressed by
$\kappa(\boldsymbol{R},\boldsymbol{r}) \equiv \kappa(\boldsymbol{R},\boldsymbol{r};\sigma,-\sigma)$,
where  $\boldsymbol{r}=\boldsymbol{r_1}-\boldsymbol{r_2}$  is the relative distance between the correlated
nucleons and $\boldsymbol{R}=(\boldsymbol{r_1}+\boldsymbol{r_2})/2$
is the center of mass coordinate. The function  $\kappa(\boldsymbol{R},\boldsymbol{r})$ calculated
at $\boldsymbol{r}=0$ is, up to normalization, the microscopic counterpart of Ginzburg-Landau order
parameter. This is one of the reasons why the pairing tensor in coordinate  representation is sometimes
called the wave function of the condensate. It is however worth mentioning  that the pairing tensor in
coordinate space is not the standard  wave function of two nucleons in the nucleus, normalized to unity. 
In fact,  the norm $\mathbb{N}_{\kappa}=\int |\kappa(\boldsymbol{R},\boldsymbol{r})|^2  d\boldsymbol{R} d\boldsymbol{r}$
has a well-defined meaning, i.e., it provides the probability  to find in the system two nucleons with opposite spins, 
irrespective to their spatial localization. 

In infinite uniform systems the spatial range of pairing correlations is characterized by the coherence
length. In nuclei this quantity is usually defined by (e.g., see \cite{pillet07})
\begin{equation}
\xi(\boldsymbol{R})=\frac
{(\int \boldsymbol{r}^2 |\kappa(\boldsymbol{R},\boldsymbol{r})|^2  d\boldsymbol{r})^{1/2}}
{(\int |\kappa(\boldsymbol{R},\boldsymbol{r})|^2 d\boldsymbol{r})^{1/2}}.
\label{xi}
\end{equation}
According to its definition, this quantity measures the  relative distance between the correlated
nucleons when their center of mass is located at $\boldsymbol{R}$. When in Eq. (7) the integrals
are performed also over $\boldsymbol{R}$, one obtains the average coherence length, which is
the root mean square (rms) radius of the pairing tensor $\kappa(\boldsymbol{R},\boldsymbol{r})$.
In Eq. (7) the  denominator was introduced in  analogy to  the definition of rms radius of a 
standard two-body wave function.

The BCS state (\ref{bcs}) is a superposition of pair condensates with various
pair numbers and, as such, does not  conserve  the particle number. An alternative  approach,
which conserves the particle number, is the particle-number projected-BCS (PBCS) model 
based on the trial state
\begin{equation}
|PBCS(N)> = (\Gamma^+)^{N/2} |0>,
\end{equation} 
where $\Gamma^+=\sum_k  y_k a^+_k a^+_{\bar{k}}$. In the simplest approximation, called projection after variation method, 
the pair operator $\Gamma^+$ is calculated with the mixing amplitudes extracted from BCS calculations, that is $y_k=v_k/u_k$. 
In the variation after the projection method, employed in this study, the amplitudes $y_k$ are determined from the minimization
of $<PBCS|H|PBCS>$ and imposing the normalization condition for the PBCS state.

In the PBCS approximation the quantity which plays the role of the pairing tensor is
\begin{equation}
\kappa(x_1,x_2) \equiv \bra{PBCS(N)}\hat{\psi}(x_2)\hat{\psi}(x_1)\ket{PBCS(N-2)}.
\label{kp}
\end{equation}
The localization properties of the pairing tensor in the PBCS approximation, discussed in
the next section, will allow to estimate how much are affected the two-body correlations 
by the particle number fluctuation.

In the coordinate representation the  PBCS state can be written as
\begin{equation}
\Psi(x_1,x_2,..,x_N)= \mathcal{A} \{\phi(x_1,x_2) \phi(x_3,x_4) ...\phi(x_{N-1},x_N)\} ,
\label{psi}
\end{equation}
where $\phi(x,x')$ is the two-body wave function corresponding to the Cooper pair operator $\Gamma^+$,
which is defined by
\begin{equation}
\phi(x,x') = \sum_k y_k f_k (x) f_{\bar{k}}(x').
\label{cooperwf}
\end{equation} 

The BCS state can be written as a superposition of functions (\ref{psi}) with various numbers of pairs. 
In this case the two-body wave  function $\phi(x,x')$ has the expression
\begin{equation}
\phi(x,x') = \sum_k \frac{v_k}{u_k} f_k (x) f_{\bar{k}}(x')
\label{bcsfunc}
\end{equation}
where $v_k$ and $u_k$ are the  variational parameters of the BCS state.

In  analogy to Eq. (7), we associate to  the wave functions 
\eqref{cooperwf} and \eqref{bcsfunc} a coherence length defined by 
\begin{equation}
\xi_C(\boldsymbol{R})=\frac
{(\int \boldsymbol{r}^2 |\phi(\boldsymbol{R},\boldsymbol{r})|^2  d\boldsymbol{r})^{1/2}}
{(\int |\phi(\boldsymbol{R},\boldsymbol{r})|^2 d\boldsymbol{r})^{1/2}}.
\label{xic}
\end{equation}
The rms radius of the wave function $\phi$ is denoted  by $<\xi_C>$.

The two-body function $\phi(x,x')$  is called below the wave function of the Cooper pair. Its physical meaning is different from
the two-body correlation function $\kappa(x,x')$. As mentioned above, the latter is related to the amplitude for two-particle
transfer and therefore can be eventually probed in two-nucleon transfer reactions. On the other hand, the Cooper pair 
wave function cannot be probed directly. However, its localization properties are fundamental because they determine 
in what coupling regime is a many-body system. Thus, if the size of the Cooper pair is larger than the average
distance between the particles, the system is in the weak coupling BCS regime. On the other extreme, if the size of
the Cooper pair is smaller than the mean distance between the particles the system is in the 
Bose-Einstein condensation (BEC) regime.  It should be noted that in the BEC limit, the Cooper pair wave function and the pairing tensor in
coordinate space become similar \cite{ortiz}. 

In this study the localization properties of  the pairing tensor and the Cooper pair wave function are calculated with a
mean field described in the RMF approximation. In RMF the nucleons are represented by Dirac spinors and interact
through  the meson fields $\sigma$, $\rho$, and $\omega$ . The nucleons and the mesons are described by the  Lagrangian 
\begin{multline}
\mathcal{L}= \bar{\psi} \big[ i \gamma^{\mu}\partial_{\mu} - M - g_{\sigma}\sigma - g_{\omega} \gamma_{\mu}\omega^{\mu} - g_{\rho} \gamma_{\mu} \vec{\rho}\cdot\vec{\tau}^{\mu} - \\ \frac{f_\pi}{m_\pi} \gamma_5\gamma_{\mu}\partial^{\mu}\vec{\pi}\cdot\vec{\tau} - e \gamma_{\mu} A^{\mu} \left( \dfrac{1-\tau_3}{2}\right)\big] \psi  + \mathcal{L}_k,
\label{lagrangian}
\end{multline}
where $\psi$ stands for the Dirac spinor describing the nucleon, $\vec{\tau}$ denotes the Pauli matrices, $\pi$ is the pion field and
$A^\mu$ represents the electromagnetic field.  The Hamiltonian of the system is  obtained from \eqref{lagrangian} through a Legendre transformation and treated in the Hartree approximation (e.g., see \cite{tam} and the references quoted  therein). For the coupling constants 
which define the relativistic mean field we employed the DD-ME2 parametrization \cite{dd-me2}.

The pairing correlations are described  in the RHB and RMF+PBCS approximations while as pairing force we employ
the Gogny interaction D1S \cite{d1s}.  In the RHB calculations the pairing force is treated within the general Bogoliubov approach. 
After the convergence of RHB calculations we determine the canonical basis, in which the  Bogoliubov 
equations take the  BCS form. Then, with the occupation probabilities and single-particle wave functions
corresponding to the canonical basis we compute the pairing tensor (6) and the Cooper pair wave
function (12).  Finally, the two-body correlations  functions  are  expressed in the relative and the center 
of mass coordinates using the Brody-Moshinsky transformation (see Ref. \cite{pillet07} for more details).
In the RMF+PBCS calculations
the PBCS equitations  are solved in the variation after projection approach using the recurrence relation method \cite{sandulescu_errea}.
Both the RMF+PBCS and RHB calculations are performed in an axially-deformed oscillator basis composed by 12 shells.

\section{Results}

To illustrate the localization  properties of pairing correlations  we present the results for three 
open shell nuclei, namely $^{66}$Ni, $^{124}$Sn and $^{200}$Pb.  The binding energies, the rms neutron
radii and the neutron pairing energies provided by  RHB and RMF+PBCS calculations
are displayed in Table I.  One can observe that, compared to RHB,
the RMF+PBCS calculations give larger pairing and binding energies and smaller neutron radii. Since the
scope of this paper is not to analyze these global quantities, we just point to the fact  that the binding energies
are rather close to the experimental values. This shows that the relativistic models employed in this paper 
are realistic enough for the study of  the two-body correlations discussed below.

\begin{table*}
\begin{tabular}{|l|ccc|cc|cc|}
    \hline
    Nuclei &\multicolumn{3}{c}{E(MeV)}& \multicolumn{2}{|c|}{$\text{r}_{\text{n}}$(fm)}  & \multicolumn{2}{c|}{E$_{\text{pair}}$(MeV))}\\
    \hline
    \quad &RHB &RMF+PBCS & Exp &RHB &RMF+BCS &RHB &RMF+BCS \\
    ${}^{66}$Ni & 575.6 & 576.9 & 576.81 & 3.92 & 3.89 & 5.14 & 6.21 \\
    ${}^{124}$Sn & 1048.7 & 1050.2 & 1049.96 & 4.81 & 4.77 & 8.9 & 11.5 \\
    ${}^{200}$Pb & 1574.7 & 1578.2 & 1576.37 & 5.56 & 5.42 & 13.4 & 15.7 \\
    \hline
    
\end{tabular}
 \caption{Ground state energies ($E$), neutron rms radii ($\text{r}_{\text{n}}$) and pairing energies ($E_{pair}$)
 calculated in RHB and 
 RMB+PBCS approximations.  In the column  "Exp" are given the experimental ground state energies \cite{exp}.} 
 \label{table:prop}
\end{table*}

We start by presenting the global properties of the two-body correlations functions introduced in the previous
section. In Table II are given the average coherence lengths associated to the pairing tensor and the Cooper
pair wave function  calculated in RHB and RMF+PBCS approximations. In the same table is also given the
mean distance  $<d_n>$ between two neutrons, estimated from the  neutron density, i.e.,  
$<d_n> = 1/\rho_n^{1/3}$, where $\rho_n=\frac{N}{\frac{4}{3}\pi r_n^3}$ and $r_n$ are the rms radii
displayed in Table I. It is interesting to observe that, contrary to the general believe, the average coherence 
lengths  are smaller than the size (i.e., the diameter) of the nucleus. On the other hand, the average coherence
lengths are larger than the mean distance $<d_n>$ between the nucleons, which is consistent  to the fact that in nuclei
the pairing correlations are in the BCS regime. 

\begin{table*}

\begin{tabular}{|l|cc|cc|c|}
    \hline
    Nuclei &\multicolumn{2}{c|}{$<\xi>$} & \multicolumn{2}{c|}{$<\xi_C>$}  & $<d_n>$\\
    \hline
    \quad &RHB &RMF+PBCS &RHB &RMF+BCS &\quad \\
    ${}^{66}$Ni & 4.05& 4.49 & 3.90 & 4.35 &1.86\\
    ${}^{124}$Sn & 4.85& 5.69 & 4.48 & 5.16 &1.84\\
    ${}^{200}$Pb & 6.81& 7.31 & 6.33 & 6.83 &1.82\\
    \hline

\end{tabular}

\label{table2}

\caption{The average coherence lengths   corresponding to 
pairing tensor  and Cooper pair wave function.  $<d_n>$ is the mean 
distance between two neutrons.}

\end{table*}

 The spatial localization of $|\kappa(R,r)|^2$ is shown in Fig. 1. One can observe that in RHB the two-body
 correlations are concentrated mainly at small relative distances between the nucleons, a result similar to the
 previous HFB studies. It is also worth noticing that, compared to RHB, in the RMF+PBCS calculations the  localization is enhanced
  in the center of the nucleus. More details about the localization properties of the pairing tensor 
are presented in Fig. 2. The latter displays the radial dependence of $|\kappa(R,r)|^2$ for various values of the
center of mass, namely R$_{\text{cm}}=1.0$ fm, R$_{\text{cm}}=r_n$, 
and R$_{\text{cm}}=R_{min(\kappa)}$ (the value of the center of mass for which  the coherence length 
has a minimum, see Fig. 3).  One can see that in the interior of the  nucleus $|\kappa(R,r)|^2$ has an oscillating 
behavior in the relative coordinate, which is due to the superposition of single-particle states with 
various numbers of nodes.  Far out in the surface region, at R$_{\text{cm}}=R_{min(\kappa)}$, the function 
$|\kappa(R,r)|^2$ has no more oscillations and is vanishing for $r >$ 4 fm.  From Fig. 2 one can see also that 
for $^{124}$Sn and $^{200}$Pb the two-body correlations at small distances are significantly larger in 
RMF+PBCS compared  to RHB. This is related to the fact that in PBCS the occupation probability of the 
states closer to the Fermi level, in this case the states  $1h_{11/2}$  and $1i_{13/2}$,  is increased 
compared to BCS \cite{sandulescu_pbcs}.  
Since these states have no nodes in the radial variable,  they confine stronger the nucleons at smaller distances.
%tb
\begin{figure}
\includegraphics[scale=0.9]{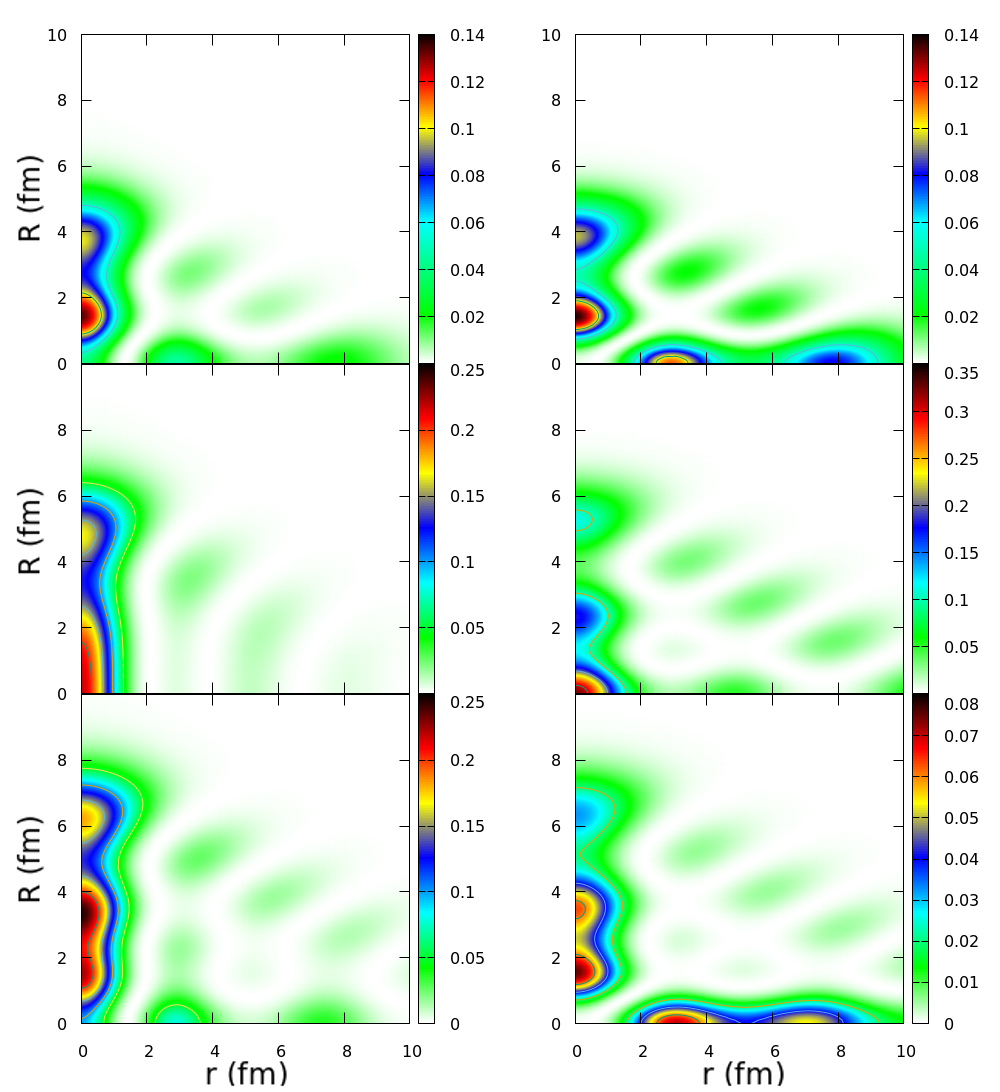}
\caption{(Color online) Two-body correlations function  $|\kappa(R,r)|^2$  provided by  RHB (left) and 
RMF+PBCS (right) calculations. From up  to down the results correspond to $^{66}$Ni, 
$^{124}$Sn and $^{200}$Pb.}
\label{comp_2d}
\end{figure}

\begin{figure}[H]
\includegraphics[scale=0.7]{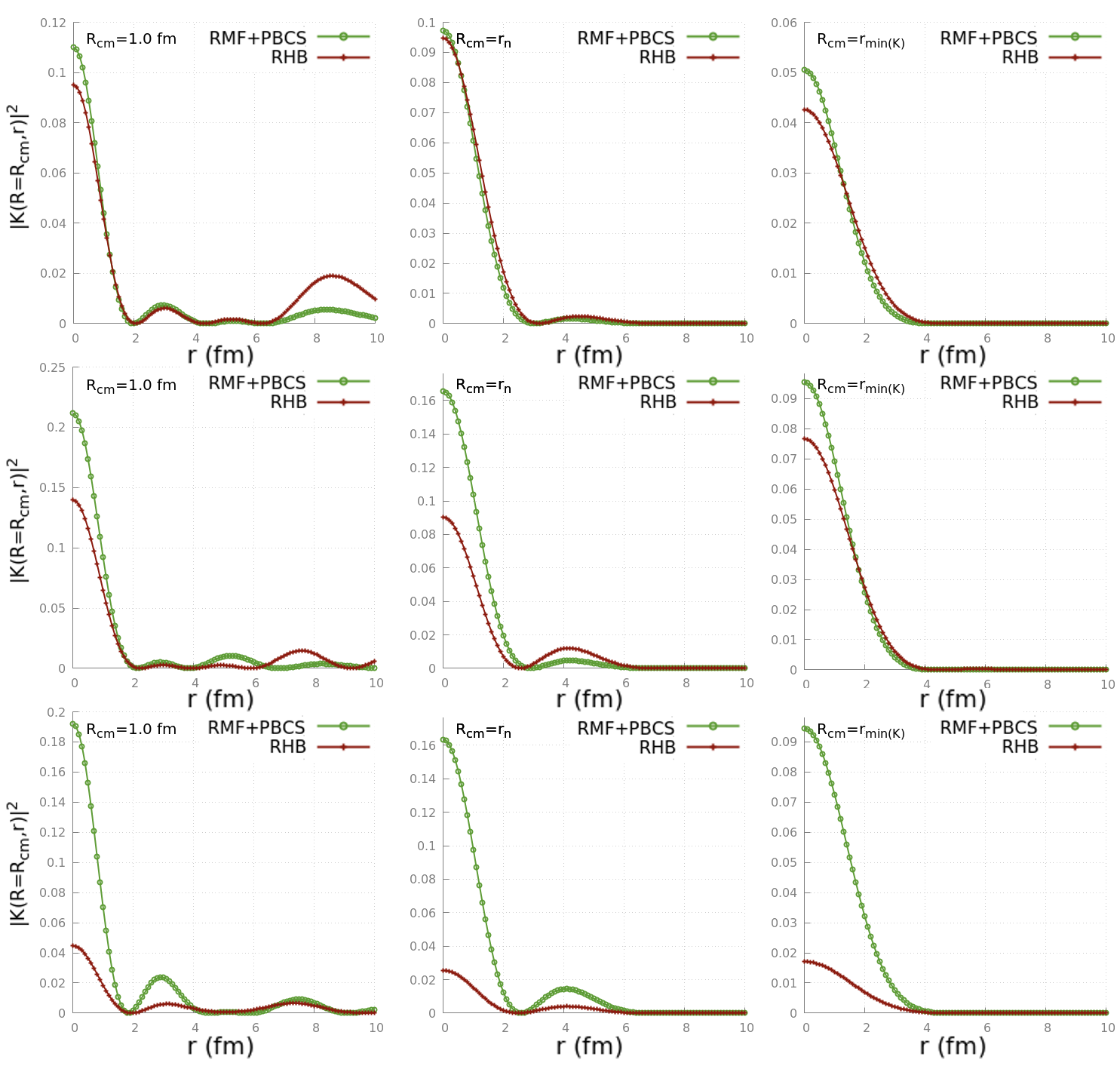}
\caption{(Color online) The dependence of $|\kappa(R,r)|^2$ on relative distance $r$ shown for R$_{\text{cm}}=1.0$ fm (left), 
R$_{\text{cm}}=r_n$ (middle) and R$_{\text{cm}}=$R$_{\text{min}(\kappa)} (right)$. From up to down 
the results correspond to  $^{66}$Ni, $^{124}$Sn and $^{200}$Pb. }
\label{cut1D}
\end{figure}

\begin{figure}
\includegraphics[scale=0.3]{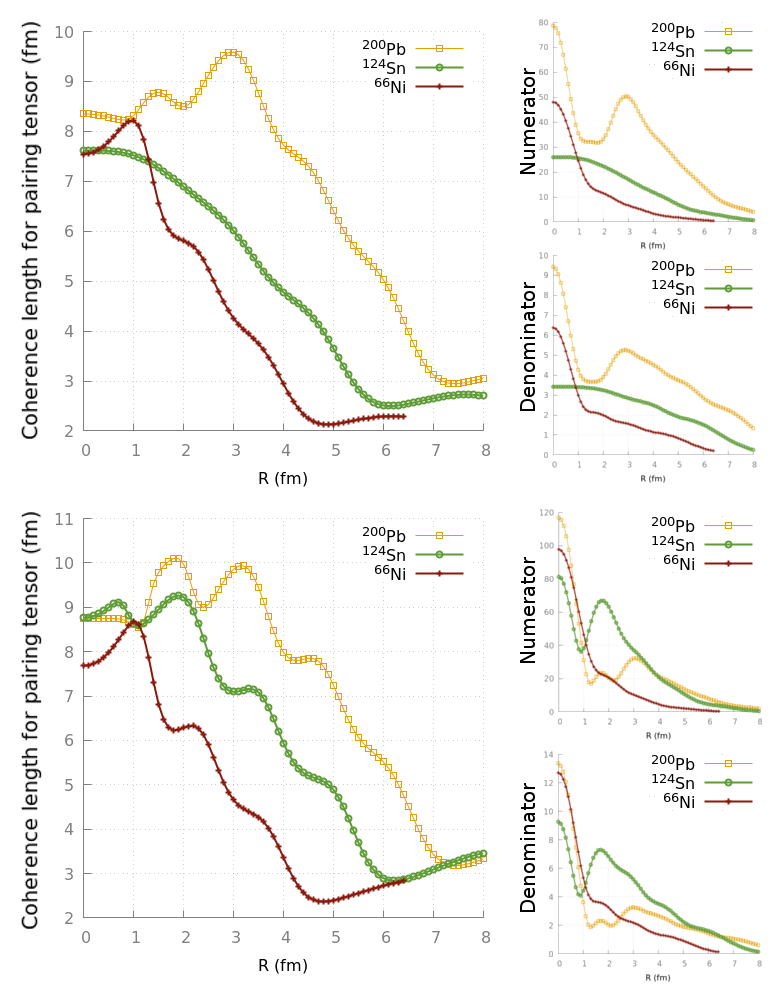}
\caption{(Color online) The coherence length corresponding to the pairing tensor
 (Eq. \ref{xi})  provided by RHB (upper panel) and RMF+PBCS
 (lower panel) calculations. The numerators and denominators of 
 Eq. (\ref{xi}) are plotted on the r.h.s.}
 \label{kappa_coh}
\end{figure}

We now analyse the coherence length $\xi(\boldsymbol{R})$ defined by Eq. (\ref{xi}).
The results obtained in RHB and RMF+PBCS calculations are displayed in  Fig.  \ref{kappa_coh}.
In the same figure are also shown the results for the numerators and the denominators of Eq. (\ref{xi}).
One  can notice that for all  nuclei the coherence length is of the order 
of $2 \times r_{n}$ in the interior of the nucleus, then decreases quite fast and reaches a minimum
far out in the nuclear surface.  As seen from Fig. \ref{kappa_coh}, the minima appear due to the
different slopes of the nominator and denominator, the latter decreasing faster at distances
larger than the neutron radius. 

\begin{figure}
\includegraphics[scale=0.4]{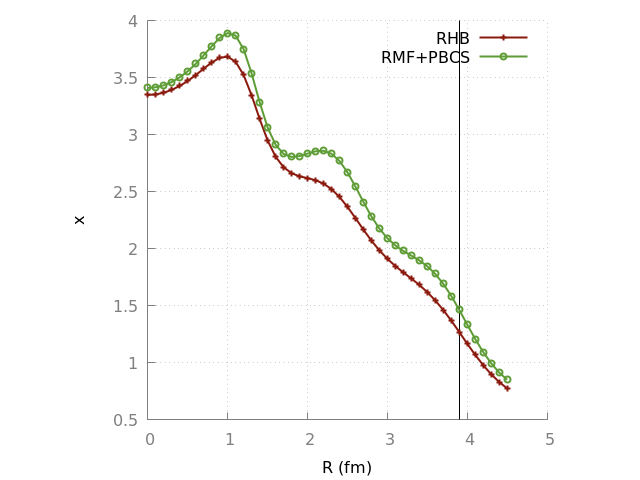}
\caption{(Color online) The ratio \eqref{xi} between the coherence length and the local distance between the neutrons for  $^{66}$Ni.
The vertical line corresponds to the neutron rms radius r$_n$.}
\label{ratio}
\end{figure}

For all nuclei the coherence length is larger 
than the local distance between the neutrons. This fact is illustrated in Fig. \ref{ratio} which shows, for $^{66}$Ni, the
ratio $X=\dfrac{\xi (\boldsymbol{R})}{d_n(\boldsymbol{R})}$, where $d_n(\boldsymbol{R})=\rho^{-1/3}_n(\boldsymbol{R})$
is the local distance between two neutrons evaluated at the neutron density  $\rho_n(\boldsymbol{R}$) .
One can observe that at $\boldsymbol{R}=r_n$ the coherence length is 
larger than the local distance between the nucleons. One can also notice that at its minimum value the coherence 
length is comparable to the local distance between the nucleons. Hence, at the minimum value of the coherence
length the two-body correlations induced by pairing are negligible. This fact can be also seen from the numerator 
of Eq. (\ref{xi}) which, as shown in Fig. \ref{kappa_coh}, is very small at the minimum value of $\xi(\boldsymbol{R})$.  

The coherence lengths shown in Fig. \ref{kappa_coh} have a similar pattern with the ones provided by the
HFB approach \cite{pillet07}. At a closer inspection one can notice that there are  
quantitative differences between the predictions of RHB and HFB but, in general, these
differences are rather small, as seen, for example, for the nucleus $^{66}$Ni. 
The largest differences are expected for the nuclei in which RHB and HFB predicts very
different mean fields and spin-orbit splittings. These are the nuclei close to the neutron drip line, 
especially the ones with a state of low angular momentum in the vicinity of the Fermi level.
An example of such a nucleus, considered in Ref.\cite{pillet07}, is $^{84}$Ni. In the HFB approach the
coherence length of this nucleus is much larger  than for $^{66}$Ni (see Fig. 5 of Ref. \cite{pillet07}). 
We have found that this is not the case in the RHB calculations. The main reason is related to the confinement
of the state $3s_{1/2}$. Thus, in HFB this state is very little bound and therefore its large spatial extension
affects significantly the neutron skin and the coherence length. This is not happening in the RHB calculations because
the relativistic mean field is deeper and therefore the state $3s_{1/2}$ is much more  bound than 
in HFB.

\begin{figure}
\includegraphics[scale=0.45]{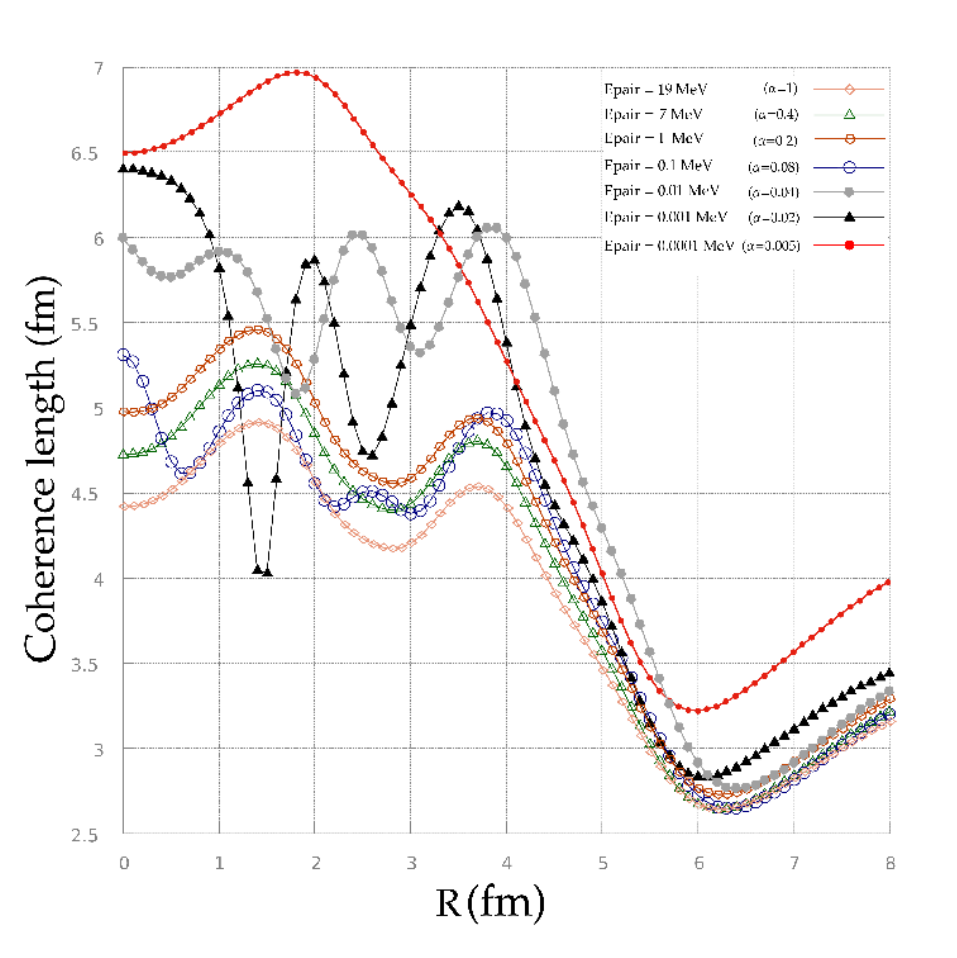}
\caption{(Color online) The coherence length for $^{120}$Sn calculated in RMF+PBCS approach for various intensities of the pairing force, scaled by the factors $\alpha$. For each scaling
factor is shown the pairing energy. }
\label{cl_120sn}
\end{figure}

In what follows we analyse the dependence of the coherence length on the intensity of the pairing force.
This issue was discussed in Ref. \cite{pillet07} in the framework of HFB for the particular case of
the nucleus $^{120}$Sn. Here we perform a similar study in the RMF model and treating the pairing
force in the PBCS approach. The advantage of this approach is that one can progressively reduce  
the pairing correlations until they are vanishing, reaching the mean-field limit. It should be 
reminded that in the BCS-like calculations the pairing correlations in nuclei are 
found only above a critical value of the pairing strength. The coherence length predicted by  RMF+PBCS 
for $^{120}$Sn is shown in Fig. \ref{cl_120sn}.  
In order to see the dependence of the coherence length on pairing correlations,
in Fig. \ref{cl_120sn} are shown the results obtained by reducing the intensity of the pairing force with various scaling factors. These results are obtained by solving the PBCS equations with the single-particle states of the major shell provided by the relativistic mean field, which is kept fixed during the calculations. One can notice that the coherence length is not changing much for a quite strong reduction of the pairing strength, as found also in the
HFB calculations \cite{pillet10,pastore}. However, at a certain point, when the interaction is becoming much weaker, the coherence length 
starts increasing until it reaches the RMF limit. 

\begin{figure}
\includegraphics[scale=0.25]{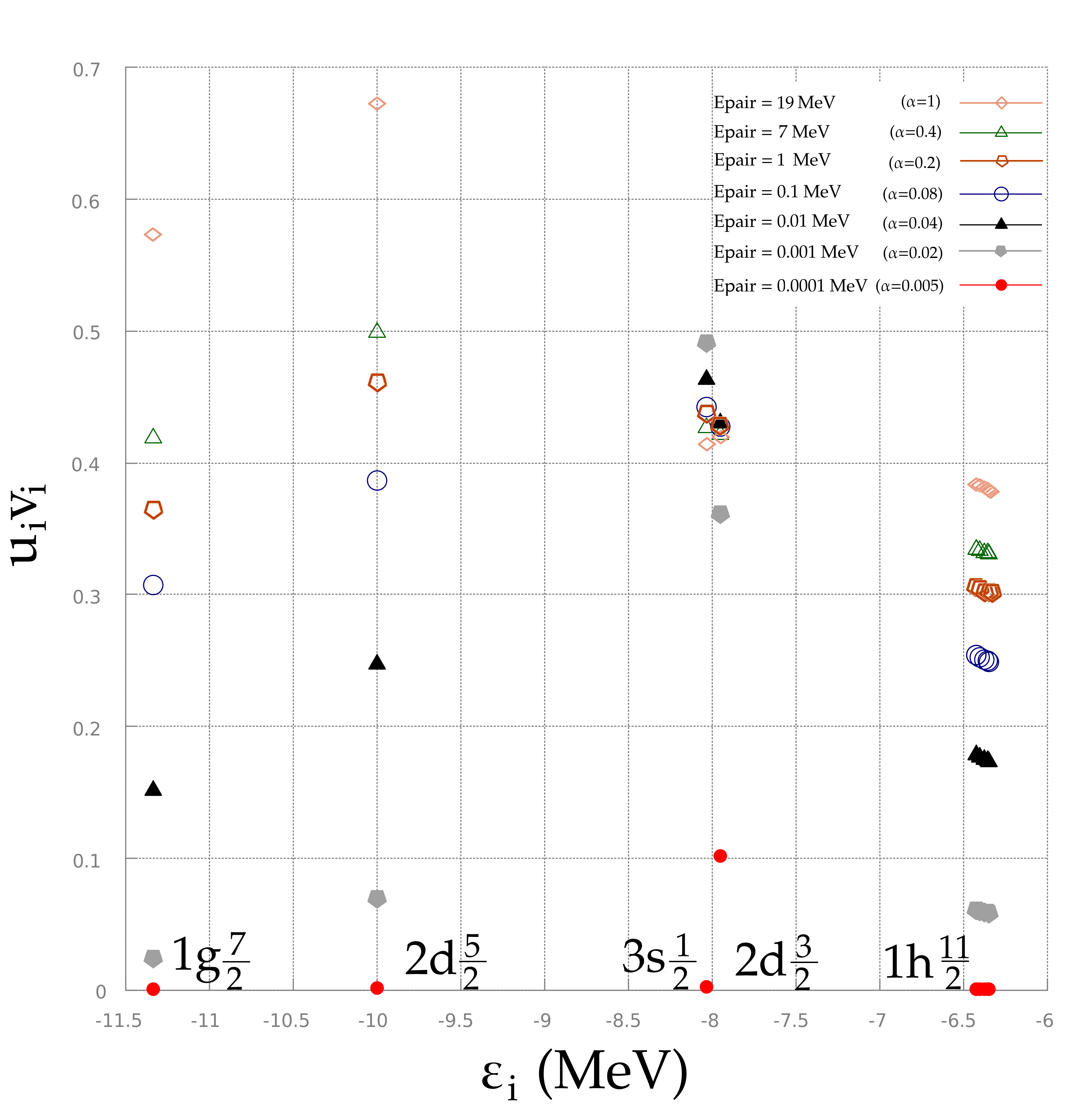}
\caption{(Color online) The quantity $k_i=u_i v_i$ for the single-particle states of
$^{120}$Sn from the major shell. The results correspond to the RMF+PBCS calculations done with the same intensities of the pairing force used in Fig. \ref{cl_120sn}. }
\label{k_i}
\end{figure}

To understand these results, we analyse how the states of the major shell contribute to the coherence length.
The contribution of a state $i$ to the the coherence length  depends on 
$k_i=<PBCS(N-2)|c_i c_{\bar{i}}|PBCS(N)>$. 
In order to simplify the analysis, we approximate this quantity by $k_i=u_iv_i$, where $v^2_i$ is the occupation
probability of the state $i$ provided by the PBCS calculations and $u^2_i=1-v^2_i$. In the BCS approach $k_i$  
represents the pairing tensor in the configuration representation. The dependence of $k_i$ on the pairing strength
is shown in Fig. \ref{k_i}. From this figure one can see that in the RMF limit the coherence length is 
built on the
orbital $2d_{3/2}$. Since this state is fully occupied in the RMF limit, the coherence length 
in this limit represents in fact the spatial correlations induced by the localisation properties of the single-particle wave function $2d_{3/2}$, and not by pairing.

The effect of pairing on coherence length  is produced through the mixing of various
single-particle states, each of them coming with their intrinsic localisation properties and contributing
according to their pairing tensor $k_i$. Respecting to the localisation properties of single-particle
states of the major shell, in Ref .\cite{pillet10} it was shown that their intrinsic coherence length, 
calculated by applying Eq. (7) for two nucleons coupled to $J=0$ and sitting in a given orbital, 
have in average a similar shape (see Fig. 17 of \cite{pillet10}). A rather different shape has the coherence length for $3s_{1/2}$. 
This state, having a small degeneracy, will have a small contribution to the total coherence length 
when the mixing with the other states of the major shell, of  higher degeneracy, is significant.
From Fig. \ref{k_i} one can observe that when the pairing strength is switched on, the first levels which mix up are the states $2d_{3/2}$ and $3s_{1/2}$. This mixing has a significant
effect on the total coherence length because these states have quite different localisation properties and because their $k_i$ values are changing 
differently with the strength of the force, as seen in Fig. \ref{k_i}.
When the strength of the force is increasing enough, the 
contributions of the other three states of the major shell come into play.  Since the degeneracy of these states is high, 
their contributions to the coherence length become quickly dominant. As seen in Fig. \ref{cl_120sn}, once the effect of these states 
is taken into account, a further increase of the pairing strength does not affect much the coherence length. 
The main reason is that the $k_i$ values of these three states are increasing proportionally to each other
and therefore, due to the denominator of Eq. (\ref{xi}), this increase is largely canceled out. Moreover,
since these states have similar localisation properties, a redistribution of their contribution in Eq. (\ref{xi}), 
due to the increase of pairing, is not expected to affect much the total coherence length. 

How important are the pairing correlations for the coherence length can be appreciated by comparing the result for the physical value of the pairing force with the result for the RMF limit. One can thus observe that the pairing correlations
have a significant contribution to the coherence length, both in the bulk and in the region of the minimum. Indeed, due to pairing, the coherence
length is decreasing in the bulk from about 6.5 fm to 4.5 fm while the minimum of the coherence length is shifted down from about 3.3 fm to 2.6 fm. Finally on this issue we would like
to emphasize that the weak dependence of the coherence length on the intensity of pairing force, especially in the minimum region, observed in Refs. \cite{pillet10,pastore}
and in the present study, is the consequence of the cancelations discussed above and therefore should be not interpreted as an indication that in nuclei the coherence length is ruled just
by finite-size effects.

\begin{figure}
\includegraphics[scale=0.3]{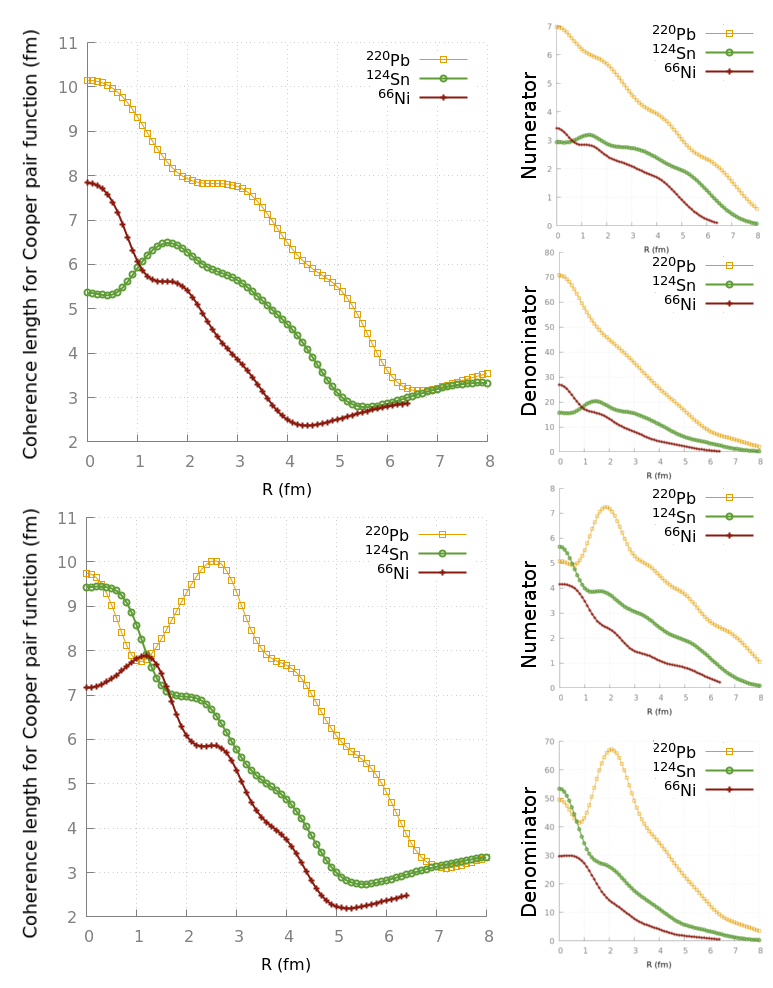}
\caption{(Color online) The coherence length corresponding to the Cooper pair wave function
 (Eq. \ref{xic}) provided by RHB (upper panel) and RMF+PBCS
 (lower panel) calculations. The numerators and denominators of Eq. (\ref{xic}) are plotted
 on the r.h.s.}
 \label{kappa_coh2}
\end{figure}

Fig. \ref{kappa_coh2} shows the coherence length associated to the Cooper pair wave function, defined by Eq. (\ref{xic}). 
One can observe that the coherence lengths associated to the Cooper pair and the pairing tensor
have a similar behavior, especially in the RMF+BCS case. The reason of this similarity comes from the fact that both quantities measure the spin correlations between 
two nucleons. However, the physical meaning of the two quantities is different because the pairing tensor
is referring to the correlations of two generic neutrons in the nucleus while  $|\phi(R,r)|^2$ is  related to the
correlations of the  neutrons  inside the Cooper pair.
This difference should be kept in mind when are 
drawn conclusions from the results presented in Fig. \ref{kappa_coh2}. 
Thus, from Fig. \ref{kappa_coh2}  one cannot draw the conclusion that in the nucleus there are Cooper 
pairs of small size in the surface and Cooper pairs of large size in the interior of the nucleus. 
This is because in BCS-type models, as seen in Eq. (10),  there is only one  Cooper pair wave function, 
identical for all the paired nucleons, which is defined globally by all values of the center of mass and
relative coordinates.

\section{Summary}
In this study we have investigated the properties of the coherence length associated to the
pairing tensor in the framework of RHB and RMF+PBCS models. Thus, taken as examples the nuclei
$^{66}$Ni, $^{124}$Sn and $^{200}$Pb, we have  shown
that the coherence length predicted by the relativistic models has a pattern which is 
similar to what was found earlier in the HFB calculations: is maximum in the bulk and
then is decreasing to a minimum of the order of 2.5-3 fm out in the surface.   

In the RMF+PBCS approach we have also analysed, for the nucleus $^{120}$Sn, the dependence
of the coherence length on the intensity of the pairing force. This analyses shows that the
pairing correlations is reducing the coherence length, compared to the RMF results, by
25-30$\%$, both in the bulk and in the region of the minimum.

Finally we have studied the coherence length associated to  the Cooper pair wave function
and we have shown that it has a similar shape as the coherence length corresponding to the pairing
tensor. In both RHB and RMF+PBCS calculations  the size of the Cooper pair  is much  larger than
the mean distance between the nucleons,  a fact consistent with the BCS regime of pairing correlations
in nuclei. 

\section*{Acknowledgements}
N. S. thanks the hospitality of Nuclear Theory Group of CSIC-Madrid, where the paper was written.
This work was supported by the French-Romanian agreement IN2P3-IFIN and by a grant of  Romanian Ministry of Research and
Innovation, CNCS- UEFISCDI,  project number PN-III-P4-ID-PCE-2016-3-0481.

\end{document}